# Gate-tunable diode and photovoltaic effect in an organic-2D layered material p-n junction§

Saül Vélez[a,†], David Ciudad[a,†], Joshua Island[b], Michele Buscema[b], Oihana Txoperena[a], Subir Parui[a], Gary A. Steele[b], Fèlix Casanova[a,c], Herre S. J. van der Zant[b], Andres Castellanos-Gomez[b,d,*], Luis E. Hueso[a,c,*]

[a] CIC nanoGUNE, 20018 Donostia-San Sebastian, Basque Country, Spain

[b] Kavli Institute of Nanoscience, Delft University of Technology, Lorentzweg 1, 2628 CJ Delft, The Netherlands

[c] IKERBASQUE, Basque Foundation for Science, 48011 Bilbao, Basque Country, Spain

[d] Instituto Madrileño de Estudios Avanzados en Nanociencia (IMDEA-Nanociencia), 28049 Madrid, Spain

[†] These authors have contributed equally to this work

* E-mail: (A.C.-G.) andres.castellanos@imdea.org; (L.E.H.) l.hueso@nanogune.eu.

§ Electronic supplementary information (ESI) available. See DOI: 10.1039/C5NR04083C

**Abstract**

The semiconducting p-n junction is a simple device structure with great relevance for electronic and optoelectronic applications. The successful integration of low-dimensional materials in electronic circuits has opened the way forward for producing gate-tunable p-n junctions. In that context, here we present an organic (Cu-phthalocyanine)–2D layered material ($MoS_2$) hybrid p-n junction with both gate-tunable diode characteristics and photovoltaic effect. Our proof-of-principle devices show multifunctional properties with diode rectifying factors of up to $10^4$, while under light exposure they exhibit photoresponse with a measured external quantum efficiency of ~ 11%. As for their photovoltaic properties, we found open circuit voltages of up to 0.6 V and optical-to-electrical power conversion efficiency of 0.7 %. The extended catalogue of known organic semiconductors and two-dimensional materials offer the prospect for tailoring the properties and the performance of the resulting devices, making organic-2D p-n junctions promising candidates for future technological applications.

**Introduction**

The p-n junction is one of the basic building blocks in electronics since it has a vast range of applications such as rectifying diodes, optical sensors, light emitters and photovoltaic solar cells[1]. P-n junctions are traditionally realized by putting two semiconductors with opposite doping in close contact, which leads to a built-in electric field at the junction. In bulk semiconductors, the charge depletion region is rather wide and fixed by the chemical doping and the band structures of their constituents. The charge density in semiconductors can be altered via electrostatic gating, providing a powerful tool to control material properties and the operation of diverse electronic devices, including p-n junctions[2]. However, in typical vertical semiconductor structures the thickness of the materials screens the electrostatic field, which hinders the control with gate voltages and limits the performance of such devices. This limitation can be overcome by using ultra-thin two-dimensional (2D) layered materials[3–6]. Atomically thin semiconductors do not fully screen the electric field, allowing electrostatic control over both the junction and the first atomic layers of the adjacent material[7–9]. P-n junctions based on vertically-stacked 2D layered materials have recently demonstrated gate controllable rectifying characteristics, large photovoltaic effect and electroluminescence[10–15]. Despite of the emergent interest in combining different 2D materials due to their potential in device engineering[16,17], the studies focused on the combination of atomically thin semiconductors with other functional materials, in particular to produce p-n junctions, are still scarce[7,18–20].

Here we present a new heterostructure which links a p-type organic material with an atomically thin n-type inorganic semiconductor in a vertical p-n junction configuration. This platform allows to exploit the advantages offered by 2D materials together with the excellent optoelectronic properties, flexibility, large area and low-cost production offered by organic materials[21,22], in a gate-controllable configuration. In addition, the natural advantage of integrating 2D layered materials in vertical p-n junction devices provides the fabrication of atomically flat interfaces, a good electrostatic control over the junction, and a great potential for down scalability. As a proof of concept, we employ p-type Cu-phthalocyanine (CuPc), a material commonly used in organic photovoltaic cells as electron-donor material[23–25]. This molecule was chosen because of its chemical stability in ambient conditions, its optical properties and its relatively high carrier mobility[26]. As an n-type 2D semiconducting material we employ the transition metal dichalcogenide (TMD) $MoS_2$, a material which is attracting wide attention for its excellent electronic and optoelectronic properties[6,27–34]. Recent developments in large area synthesis[35] could also offer a way to produce $MoS_2$ devices at relatively low cost with large area coverage and scalability.

Our CuPc-$MoS_2$ devices exhibit different gate-controlled integrated functionalities, including rectification as well as photoresponse with both photogating and photovoltaic effect under light exposure, demonstrating that organic-2D p-n junction devices are potential candidates to be applied as diode-rectifiers, light sensors and solar cells.

**Experimental**

**Devices Fabrication**

The fabrication steps followed to produce a CuPc-MoS$_2$ vertical p-n junction are summarized in Fig. 1a. Briefly, large MoS$_2$ flakes are placed by a PDMS dry transfer technique[36] onto a Si$^{++}$/SiO$_2$ substrate, then e-beam lithography and Ti (4 nm)/Au (30nm) metal deposition (and lift-off) are used to define the electrodes. A second e-beam lithography step is used to mask the substrate with PMMA and to open a window, while finally CuPc (30 nm-thick layer) is thermally evaporated to produce the junction. A detailed description of the fabrication of the devices can be found in Materials and methods section. A sketch of a final device and the electric measurement scheme is shown in Fig. 1b. Fig. 1c and 1d show optical images of a bilayer MoS$_2$ flake before and after the device fabrication, respectively. In the same device, the field-effect characteristics of the MoS$_2$ flake can also be measured. CuPc field-effect transistors were fabricated simultaneously on a different part of the wafer to characterize the electrical properties of the organic film without any spurious screening effects. Several CuPc-MoS$_2$ devices were fabricated and characterised using either monolayer or bilayer MoS$_2$ flakes (three of each type). Qualitatively similar electronic and optoelectronic properties were found in all of them. Here we will present data corresponding to three bilayer-MoS$_2$-based devices (one sample for each set of data shown in Figs. 2-5, respectively). Representative data regarding monolayer-MoS$_2$-based devices can be found in the ESI§.

**Results and Discussion**

**Band structure of the CuPc-MoS$_2$ devices**

The approximate band structure of the materials and the work functions of the metals used are depicted in Fig. 1e in the idealized rigid-band approximation. The lowest unoccupied molecular orbital (LUMO) of CuPc is 3.5 eV, whereas its highest occupied molecular orbital (HOMO) level is located at 5.2 eV (ref. [25]). On the other side, the electron affinity of a monolayer MoS$_2$ is estimated to be 4.2 eV (ref. [37]). With a bandgap of 1.8 eV (1.6 eV for a bilayer; ref. [6]), its valence band is far below the HOMO level of the CuPc. Regarding the electrodes, specific metals were chosen to minimize any eventual Schottky barrier with the semiconducting materials. Ti/Au metal contacts were employed since their workfunctions, 4.3 eV for Ti and 5.3 eV for Au, lie close to the conduction band of the MoS$_2$ and the HOMO level of the CuPc, respectively[38,39]. Note that, with the fabrication process used, Ti is first deposited on top of the MoS$_2$ flake, whereas the CuPc layer is deposited on top of Au (see the ESI§ Section 1 for the metal-semiconductor contacts characterization). Charge transport will be therefore dominated by the HOMO level of the CuPc and the conduction band of the MoS$_2$ layer. Since both materials also have a large bandgap, an efficient charge separation is expected at their junction.

**Electronic characterization of the devices**

We first explore the electronic characteristics of our devices. The measurements were performed at room temperature, in vacuum and in dark conditions. Fig. 2a shows, in the same plot, the source-drain current as a function of the gate voltage, $I_{ds}(V_g)$ or transfer curve, of a representative CuPc-MoS$_2$ device (red curve), a MoS$_2$ bilayer transistor produced in the same device (blue curve) and a 30-nm-thick CuPc transistor (black curve). Since MoS$_2$ (CuPc) is a n-type (p-type) semiconductor, its electrical conduction is suppressed at large negative (positive) gate voltages, while for positive (negative) gate voltages, the MoS$_2$ (CuPc) layer becomes conductive with observed $I_{on}/I_{off}$ ratios of $10^5$. The electrical transport of the CuPc-MoS$_2$ heterojunction reflects the addition in series of the resistance of each material (ESI§ Section 2), with a sizeable conductivity for intermediate gate voltages (|$V_g$| ≤ 30 V) and measured $I_{on}/I_{off}$ ratios approaching $10^4$.

**Diode characteristics of the CuPc-MoS$_2$ devices**

Fig. 2b shows the transfer curves of the CuPc-MoS$_2$ device at different source-drain voltages. The characteristic shape of the transfer curve is conserved for all $V_{ds}$, with an asymmetry that arises from the small subthreshold swing ($V_g$ change needed for increasing $I_{ds}$ in one order of magnitude) of the MoS$_2$ layer[27]. Similar curves taken in a monolayer-MoS$_2$-based p-n junction device can be found in the ESI§ Section 3. In all cases, an efficient gate control over the heterostructure is observed. This is directly related to the use of thin MoS$_2$ flakes (either monolayer or bilayer), which allows the electric field produced by the gate to extend into the CuPc layer.

Fig. 2c shows the source-drain current *vs* voltage $I_{ds}(V_{ds})$ curves at different $V_g$, demonstrating that the device can also operate as a rectifying diode when a negative bias is applied. Current is suppressed at reverse bias whereas an exponential increase is observed at forward bias, with maximum measured forward-to-reverse current ratios approaching $10^4$, which are suitable values for logic applications. The rectifying capability of the CuPc-MoS$_2$ junction has been further tested in the presence of an applied AC source-drain voltage, showing good in-phase rectification (see inset of Fig. 2c).

The characteristics of the CuPc-MoS$_2$ diode can be further assessed by fitting the $I_{ds}(V_{ds})$ curves to the Shockley equation[40] with series and shunt resistances[41] (see the ESI§ Section 4 for more details). This fit allows the determination of the diode ideality factor ($n$), which provides information about the dynamics of the charge transfer at the p-n interface. The fit is found to be strongly dependent on $V_g$, with the best ideality factors obtained for $V_g$ values around the maxima of the $I_{ds}(V_g)$ curve. For $V_g$ = -20 V, $n$ ~ 2.9 is found, a value that increases for $V_g$ increasing (see Table S1 in the ESI§). These values indicate that the current is strongly limited by charge recombination –presumably due to the presence of trap states– rather than diffusion[42] (which would otherwise lead to $n$ = 1). Particularly, the increase in the ideality factor when $V_g$ increases correlates with the decrease in conductivity of the CuPc layer (Fig. 2a). Also from the fittings, a series resistance larger than 0.2 TΩ is found in all cases, suggesting that the main limitation in our devices is related to the lateral CuPc channel used. Finally, a shunt resistance of the order of 10-50

GΩ is obtained. We anticipate that the performance of the organic-inorganic devices could be improved by both a more controlled CuPc growth on the MoS$_2$ layer[24,43] and by reducing the length of the organic channel.

**Photogating effect in the CuPc-MoS$_2$ devices**

We turn now our attention to the optoelectronic properties of the CuPc-MoS$_2$ devices. Fig. 3 shows the transfer curve $I_{ds}(V_g)$, both in dark conditions and under illumination with white light (see Materials and methods), for a representative device operating in the forward bias ($V_{ds}$ > 0). Light absorption by the device results in both an increase in the maximum current as well as in a shift (negative in $V_g$ values) of the transfer curve. These observations suggest n-doped photogating of the MoS$_2$ layer[11,44] as the dominant effect for the photoresponse measured (see the ESI§ Section 5). The photogating effect, together with the sharp subthreshold swing characteristics of the junction, leads to a considerable generated photocurrent $I_{ph} = I_l - I_d$ and measured ratios $I_l/I_d$ exceeding one order of magnitude at the vicinity of activation threshold of the MoS$_2$ (see Fig. 3 and ESI§ Section 6), where $I_l$ and $I_d$ correspond to the current measured under illumination and in dark conditions, respectively. These features might make these devices suitable for the development of light sensors operating in the forward bias regime, as the charge current collected is very significant.

Our physical picture of the origin of the photogating is the following. The light is mainly absorbed by the 30-nm-thick CuPc layer, which is on top of the MoS$_2$, generating electron-hole pairs. The electrons diffuse towards the CuPc-MoS$_2$ junction and lead to a charge transfer from the LUMO level of the CuPc to the conduction band of the MoS$_2$. Some of the charges may be captured in long-lived trap states in either the interface itself or at defects close to the junction, effectively modifying the doping of the MoS$_2$ layer, which shifts its threshold to negative $V_g$ values. A smaller contribution may arise from the MoS$_2$ layer –smaller since it is effectively buried by the CuPc layer and has a limited light absorption due to the reduced thickness[45]– leading to charge generation under light exposure and hole transfer from the valence band of the $MoS_2$ to the HOMO level of the CuPc.

**Photovoltaic effect in the CuPc-$MoS_2$ devices**

Next, we focus on the photovoltaic characteristics of our CuPc-$MoS_2$ p-n devices. Fig. 4 shows the photovoltaic response under light exposure as a function of gate voltage (Fig. 4a and 4b), wavelength (Fig. 4c and 4d), and incident optical power (Fig. 4e and 4f) of a representative device (a dataset corresponding to a monolayer-based CuPc-$MoS_2$ device can be found in the ESI§ Section 7). The incident light on the device produces a shift of the $I_{ds}(V_{ds})$ curves (Fig. 4a, 4c and 4e), which leads to a net current at zero $V_{ds}$ (short circuit current, $I_{sc}$) and to a net voltage for null current (open circuit voltage, $V_{oc}$). These two characteristic quantities are plotted in Fig. 4b, 4d and 4f. Note that the $I_{ds}$-$V_{ds}$ curves do not show an ideal diode-like behaviour at small $V_{ds}$ (negative) values, with the current being sensitive to the applied driving potential. This behaviour is a manifestation of charge trapping and recombination effects occurring in the device in agreement with previous data reported.

Large $V_{oc}$ and $I_{sc}$ values were detected in the range of $V_g$ values where the diode has better ideality factors (see Fig. 4a and 4b), which indicates that the photovoltaic effect has its origin at the p-n junction rather than being produced, for instance, by the Schottky barriers or by photothermoelectric effects[46,47] near the contacts.

Wavelength dependent measurements were carried out across the visible spectrum (Fig. 4c and 4d). A large, fairly constant $V_{oc}$ is observed at all wavelengths, whereas the $I_{sc}$ values are correlated to the wavelength absorption spectra of the CuPc layer[48] (see ESI§ Section 8). These observations confirm that the induced photocurrent is mainly generated in the CuPc layer, while the $V_{oc}$ is related to the electronic properties of the CuPc-MoS$_2$ junction. More precisely, the $V_{oc}$ must be limited by the voltage drop between the energy levels involved in the charge transfer at the junction and charge recombination. The estimated energy difference between the HOMO level of the CuPc and the conduction band of the MoS$_2$ is $\Delta E \sim 1.0$ eV, which is indeed larger than the $V_{oc}$ values observed.

The effect of the incident optical power on the photovoltaic effect is shown in Fig. 4e and 4f. We observe a positive linear trend in the $V_{oc}$ with increasing laser optical power $P_{laser}$, with a maximum $V_{oc}$ slightly above 0.6 V. $I_{sc}$ increases in the form of a power law with $P^{\beta}_{laser}$ with an exponential factor $\beta \sim 0.35$, reflecting the role of trap charges and recombination in the optical-to-electrical power conversion[49,50]. The electrical power generated in our device ($I_{ds} \cdot V_{ds}$) under illumination at a fixed wavelength ($\lambda$ = 640 nm) is calculated from the $I_{ds}(V_{ds})$ curves shown in Fig. 4e and plotted as a function of $V_{ds}$ in Fig. 5a. As expected, the maximum of the electrical power generated, $P_{el,m}$, also increases following a power law with $P_{laser}$ in agreement with the observed $I_{sc}$ dependence. The efficiency of the device is evaluated from the ratio between $P_{el,m}$ and the incident optical power $P_{opt}$, which is plotted in Fig. 5b as a function of the incident illumination (see Materials and methods for the calculation of $P_{opt}$ from $P_{laser}$). The maximum efficiency of $\eta_{PV} = P_{el,m}/P_{opt} = 0.7$ % is obtained at the lowest incident power, and decreases as $P_{opt}$ increases. The low efficiency observed in the low-bias regime and the strong $P_{el,m}$ dependence on $P_{opt}$, which is correlated to the $I_{sc}(P_{opt})$ dependence observed, are attributed to the large series resistance and limited shunt resistance found for our CuPc-MoS$_2$ device.

Alternatively, the external quantum efficiency (EQE) –defined as the ratio of collected electron-hole pairs for incident photon– of the CuPc-MoS$_2$ devices has been evaluated at a large $V_{ds}$ (-6 V) and low incident optical powers, leading to a measured EQE of up to ~11% for an incident wavelength λ = 640 nm, laser optical power $P_{laser}$ = 0.3 μW and $V_g$ = -20 V. This suggests that there is large room for further improvement of the photovoltaic effect in heterostructure devices of this kind with reduced trap charges/recombination.

Finally, we can benchmark the performance of our multifunctional organic-2D inorganic hybrid devices against other 2D layered material based p-n junction devices. Our CuPc-MoS$_2$ devices present competitive EQE and photovoltaic effect efficiencies when compared to other multifunctional p-n photodiode devices as the ones based on single or multi-layered TMDs[10–12,15,47,51,52], devices produced by combining a TMD layer with another functional material[18,19] or few-layer black phosphorous based devices[53,54]. Besides the known strong light-matter interactions occurring in atomically thin 2D crystals[55] and its potential as a platform for photodetection[56], organic materials are also known for being capable of producing large EQE thanks to their extraordinary optical properties[57,58]. We therefore anticipate larger efficiencies in organic-2D material hybrid devices with reduced trap charges/recombination and with an optimized selection of the organic media among the available catalogue of know materials, making therefore organic-2D material junctions promising candidates for solar cell technology and energy harvesting applications.

**Conclusions**

In summary, we have shown the electronic and optoelectronic properties of a proof-of-principle organic semiconductor-inorganic 2D semiconductor p-n junction. In particular, we have produced thin CuPc-MoS$_2$ vertical devices using either monolayer or bilayer MoS$_2$ flakes and demonstrate them to be multifunctional and electrostatically controllable. These devices can operate as gate-tunable rectifying diodes with rectifying factors of up to 10$^4$ and, under optical light exposure, they exhibit photocurrent generation associated with photogating of the MoS$_2$ layer and a significantly large photovoltaic effect. Photoinduced carriers are mostly generated in the CuPc layer, while the role of the MoS$_2$ is mainly to control charge separation. As for the photovoltaic properties, we found open circuit voltages of up to $V_{oc}$ ~ 0.6 V and a maximum of 0.7 % optical-to-electrical power conversion efficiency at small bias. The EQE of the device has been tested at large $V_{ds}$ and small optical powers, leading to a maximum measured EQE ~11%. The reduction of the efficiency of the optical-to-electrical power conversion when the incident optical power increases or the applied bias is reduced is associated with resistive losses and charge recombination mainly produced in the organic layer, factors that could be improved via device engineering. The variety of already-known organic semiconductors and 2D layered materials, together with the demonstrated capabilities of our proof of concept CuPc-MoS$_2$ devices, point organic-2D material hybrids towards a promising alternative platform for future electronic and optoelectronic applications.

**Materials and methods**

**Samples fabrication**

The fabrication steps used to produce our devices are depicted in Fig. 1a. We began by transferring large MoS$_2$ layers (either monolayers or bilayers) onto a Si$^{++}$/SiO$_2$ (285nm) substrate by the PDMS dry transfer technique[36] (step 1). The flakes were first identified via optical contrast and double-checked using Atomic Force Microscopy (*Digital Instruments D3100 AFM*) and Raman spectroscopy (Renishaw in via). Two PMMA layers, 495K A4 followed by 950K A2, spin-coated at 4 krpm for 60 s and baked out 60 s at 195$^0$C, were used in steps 2 and 4. Electron-beam lithography (*Raith 150-TWO*) was used to define the electrodes (step 2) and opening a window (step 4). Two of the patterned electrodes were defined on top of the MoS$_2$ flake and a third one was left free and close to the MoS$_2$ flake. Metal deposition (step 3) was done via e-beam evaporation for Ti (4 nm) and thermal evaporation for Au (30 nm) in a vacuum chamber with a base pressure of < 10$^{-6}$ mbar. Lift-off of the electrodes was done in acetone for several hours. The CuPc layer was thermally evaporated from a Knudsen cell at a rate of 0.2 Å/s in an ultrahigh vacuum chamber with a base pressure of $10^{-9}$ mbar to fill the gap between the MoS$_2$ flake and the free electrode (step 5). In addition from using a PMMA window to define the CuPc channel, the organic deposition was carried out through a metal shadow mask to limit the organic deposition area around the region of interest. A sketch of a resulting device and the measuring scheme can be seen in Fig. 1b. A contact on the Si$^{++}$ substrate enables electrostatic gating across the SiO$_2$. Note that in our devices Ti is in contact with the MoS$_2$ flake, whereas Au is in contact with the CuPc organic layer. This allows minimizing the Schottky barrier formation at the metal-semiconductor interface and improving charge injection (or extraction).

**Electronic and Optoelectronic properties characterization**

All electronics and optoelectronics characterization was done in a *Lakeshore Cryogenics* probe station at room temperature and in vacuum (~ $10^{-5}$ mBar). Phototransistor characterization (data shown in Fig. 3) was performed using a custom Schott ACE halogen-light source emitting in the visible spectra. By accounting for optical power losses and the geometrical disposition of the source with respect to the sample, an incident optical power density of ~ 1 kW m$^{-2}$ is estimated at the surface of the device. As for the photovoltaic characterization, the light excitation was provided by diode-pumped solid-state lasers operated in continuous wave mode (CNI Lasers). The light is coupled into a multimode optical fiber (NA = 0.23) through a parabolic mirror ($f_{reflected}$ = 25.4 mm). At the end of the optical fiber, an identical parabolic mirror collimates the light. The beam is then directed into the probe station zoom lens system and then inside the cryostat. The beam was centred on the CuPc-MoS$_2$ junction, which has a spot size (on the sample surface) of 230 ± 8 μm in diameter for all wavelengths, and thus covers the whole device area.

The incident optical power density at the surface of the device is calculated by dividing the optical power of the laser $P_{laser}$ by the area of the laser spot $A_{spot}$. The absorbed optical power $P_{opt}$ is calculated by $P_{opt} = P_{laser} \cdot A_j/A_{spot}$, where $A_j$ is the area covered by the CuPc-MoS$_2$ junction. The area of the junction for the device explored in the main text (data corresponding to Fig. 4 and 5) is $A_j$ = 12.1x5 = 60.5 $\mu m^2$. Only the junction area is taken into account since this is the part were the CuPc (absorber) is in direct contact with the MoS$_2$ layer. Notice that there is not direct exposure of light to the MoS$_2$ layer in any case.

## Acknowledgements

Work supported by the European Union 7th Framework Programme under the Marie Curie Actions (256470-ITAMOSCINOM, 275176-HELIOS and 300802-STRENGTHNANO), NMP project (263104-HINTS) and the European Research Council (257654-SPINTROS), by the Spanish MINECO under Project No. MAT2012-37638 and by the Dutch organization for Fundamental Research on Matter (FOM).

## Notes and references

1    S. M. Sze and K. K. Ng, *Physics of semiconductor Devices*, Wiley, 3rd edn., 2006.

2    C. H. Ahn, M. Di Ventra, J. N. Eckstein, C. Daniel Frisbie, M. E. Gershenson, A. M. Goldman, I. H. Inoue, J. Mannhart, A. J. Millis, A. F. Morpurgo, D. Natelson and J.-M. Triscone, *Rev. Mod. Phys.*, 2006, **78**, 1185–1212.

3    K. S. Novoselov, A. K. Geim, S. V. Morozov, D. Jiang, Y. Zhang, S. V. Dubonos, I. V. Grigorieva and A. A. Firsov, *Science*, 2004, **306**, 666–669.

4    K. S. Novoselov, D. Jiang, F. Schedin, T. J. Booth, V. V. Khotkevich, S. V. Morozov and A. K. Geim, *Proc. Natl. Acad. Sci. U. S. A.*, 2005, **102**, 10451–10453.

5    K. S. Novoselov, V. I. Fal'ko, L. Colombo, P. R. Gellert, M. G. Schwab and K. Kim, *Nature*, 2012, **490**, 192–200.

6    Q. H. Wang, K. Kalantar-Zadeh, A. Kis, J. N. Coleman and M. S. Strano, *Nat. Nanotechnol.*, 2012, **7**, 699–712.

7    W. J. Yu, Z. Li, H. Zhou, Y. Chen, Y. Wang, Y. Huang and X. Duan, *Nat. Mater.*, 2013, **12**, 246–252.

8    W. J. Yu, Y. Liu, H. Zhou, A. Yin, Z. Li, Y. Huang and X. Duan, *Nat. Nanotechnol.*, 2013, **8**, 952–958.

9   S. Parui, L. Pietrobon, D. Ciudad, S. Vélez, X. Sun, F. Casanova, P. Stoliar and L. E. Hueso, *Adv. Funct. Mater.*, 2015, **25**, 2972–2979.

10  C.-H. Lee, G.-H. Lee, A. M. Van Der Zande, W. Chen, Y. Li, M. Han, X. Cui, G. Arefe, C. Nuckolls, T. F. Heinz, J. Guo, J. Hone and P. Kim, *Nat. Nanotechnol.*, 2014, **9**, 676–681.

11  M. M. Furchi, A. Pospischil, F. Libisch, J. Burgdörfer and T. Mueller, *Nano Lett.*, 2014, **14**, 4785–4791.

12  R. Cheng, D. Li, H. Zhou, C. Wang, A. Yin, S. Jiang, Y. Liu, Y. Chen, Y. Huang and X. Duan, *Nano Lett.*, 2014, **14**, 5590–5597.

13  H. Fang, C. Battaglia, C. Carraro, S. Nemsak, B. Ozdol, J. S. Kang, H. A. Bechtel, S. B. Desai, F. Kronast, A. A. Unal, G. Conti, C. Conlon, G. K. Palsson, M. C. Martin, A. M. Minor, C. S. Fadley, E. Yablonovitch, R. Maboudian and A. Javey, *Proc. Natl. Acad. Sci. U. S. A.*, 2014, **111**, 6198–6202.

14  C. O. Kim, S. Kim, D. H. Shin, S. S. Kang, J. M. Kim, C. W. Jang, S. S. Joo, J. S. Lee, J. H. Kim, S.-H. Choi and E. Hwang, *Nat. Commun.*, 2014, **5**, 3249.

15  S. Wi, H. Kim, M. Chen, H. Nam, J. Guo, E. Meyhofer, X. Liang, M. Engineering, E. Engineering, C. Science, A. Arbor and U. States, *ACS Nano*, 2014, **8**, 5270–5281.

16  A. K. Geim and I. V. Grigorieva, *Nature*, 2013, **499**, 419–425.

17  J. M. Hamm and O. Hess, *Science*, 2013, **340**, 1298–1299.

18  O. Lopez-sanchez, E. A. Llado, V. Koman, A. Fontcuberta, A. Radenovic and A. Kis, *ACS Nano*, 2014, **8**, 3042–3048.

19  D. Jariwala, V. K. Sangwan, C.-C. Wu, P. L. Prabhumirashi, M. L. Geier, T. J. Marks, L. J. Lauhon and M. C. Hersam, *Proc. Natl. Acad. Sci. U. S. A.*, 2013, **110**, 18076–18080.

20  S. Chuang, R. Kapadia, H. Fang, T. C. Chang, W. Yen, Y. Chueh and A. Javey, *Appl. Phys. Lett.*, 2013, **102**, 242101.

21  *Nat. Mater.*, 2013, **12**, 591.

22  W. Cao and J. Xue, *Energy Environ. Sci.*, 2014, 2123–2144.

23  M. G. Walter, A. B. Rudine and C. C. Wamser, *J. Porphyr. Phthalocyanines*, 2010, **14**, 759–792.

24  P. Sullivan, T. S. Jones, A. J. Ferguson and S. Heutz, *Appl. Phys. Lett.*, 2007, **91**, 233114.

25  C.-W. Chu, V. Shrotriya, G. Li and Y. Yang, *Appl. Phys. Lett.*, 2006, **88**, 153504.

26  Z. Bao, A. J. Lovinger and A. Dodabalapur, *Appl. Phys. Lett.*, 1996, **69**, 3066.

27	B. Radisavljevic, A. Radenovic, J. Brivio, V. Giacometti and A. Kis, *Nat. Nanotechnol.*, 2011, **6**, 147–150.

28	B. Radisavljevic and A. Kis, *Nat. Mater.*, 2013, **12**, 815–820.

29	A. Splendiani, L. Sun, Y. Zhang, T. Li, J. Kim, C.-Y. Chim, G. Galli and F. Wang, *Nano Lett.*, 2010, **10**, 1271–1275.

30	R. S. Sundaram, M. Engel, A. Lombardo, R. Krupke, A. C. Ferrari, P. Avouris and M. Steiner, *Nano Lett.*, 2013, **13**, 1416–1421.

31	H. Zeng, J. Dai, W. Yao, D. Xiao and X. Cui, *Nat. Nanotechnol.*, 2012, **7**, 490–493.

32	K. F. Mak, K. He, J. Shan and T. F. Heinz, *Nat. Nanotechnol.*, 2012, **7**, 494–498.

33	T. Cao, G. Wang, W. Han, H. Ye, C. Zhu, J. Shi, Q. Niu, P. Tan, E. Wang, B. Liu and J. Feng, *Nat. Commun.*, 2012, **3**, 887.

34	K. F. Mak, K. L. McGill, J. Park and P. L. McEuen, *Science*, 2014, **344**, 1489–1492.

35	K.-K. Liu, W. Zhang, Y.-H. Lee, Y.-C. Lin, M.-T. Chang, C.-Y. Su, C.-S. Chang, H. Li, Y. Shi, H. Zhang, C.-S. Lai and L.-J. Li, *Nano Lett.*, 2012, **12**, 1538–1544.

36	A. Castellanos-Gomez, M. Buscema, R. Molenaar, V. Singh, L. Janssen, H. S. J. van der Zant and G. A. Steele, *2D Mater.*, 2014, **1**, 011002.

37	S. Bertolazzi, D. Krasnozhon and A. Kis, *ACS Nano*, 2013, **7**, 3246–3252.

38	S. Das, H.-Y. Chen, A. V. Penumatcha and J. Appenzeller, *Nano Lett.*, 2013, **13**, 100–105.

39	M. Gobbi, L. Pietrobon, A. Atxabal, A. Bedoya-Pinto, X. Sun, F. Golmar, R. Llopis, F. Casanova and L. E. Hueso, *Nat. Commun.*, 2014, **5**, 4161.

40	C.-T. Sah, R. N. Noyce and W. Schockley, *Proc. IRE*, 1957, **45**, 1228–1243.

41	A. Ortiz-Conde, F. J. García Sanchez and J. Muci, *Solid State Electron.*, 2000, **44**, 1861–1864.

42	T. Banwell and A. Jayakumar, *Electron. Lett*, 2000, **36**, 291–292.

43	K. K. Okudaira, S. Hasegawa and H. Ishii, *J. Appl. Phys.*, 1999, **85**, 6453–6461.

44	O. Lopez-sanchez, D. Lembke, M. Kayci, A. Radenovic and A. Kis, *Nat. Nanotechnol.*, 2013, **8**, 497–501.

45	M. Bernardi, M. Palummo, C. Grossman and R. Scienti, *Nano Lett.*, 2013, **13**, 3664–3670.

46	M. Buscema, M. Barkelid, V. Zwiller, H. S. J. van der Zant, G. A. Steele and A. Castellanos-Gomez, *Nano Lett.*, 2013, **13**, 358–363.

47  D. J. Groenendijk, M. Buscema, G. A. Steele, R. Bratschitsch, H. S. J. van der Zant and A. Castellanos-Gomez, *Nano Lett.*, 2014, **14**, 5846–5852.

48  Q. Li, J. Yu, Y. Zhang, N. Wang and Y. Jiang, *Front. Energy*, 2012, **6**, 179–183.

49  A. Rose, *Concepts in photoconductivity and allied problems*, Interscience Publishers, New York, 1963.

50  R. H. Bube, *Photoelectronic Properties of Semiconductors*, Cambridge University Press, Cambridge, UK, 1992.

51  B. W. H. Baugher, H. O. H. Churchill, Y. Yang and P. Jarillo-Herrero, *Nat. Nanotechnol.*, 2014, **9**, 262–267.

52  A. Pospischil, M. M. Furchi and T. Mueller, *Nat. Nanotechnol.*, 2014, **9**, 257–261.

53  M. Buscema, D. J. Groenendijk, G. A. Steele, H. S. J. van der Zant and A. Castellanos-Gomez, *Nat. Commun.*, 2014, **5**, 4651.

54  Y. Deng, Z. Luo, N. J. Conrad, H. Liu, Y. Gong, S. Najmaei, P. M. Ajayan, J. Lou, X. Xu and P. D. Ye, *ACS Nano*, 2014, **8**, 8292–8299.

55  L. Britnell, R. M. Ribeiro, A. Eckmann, R. Jalil, B. D. Belle, A. Mishchenko, Y.-J. Kim, R. V. Gorbachev, T. Georgiou, S. V. Morozov, A. N. Grigorenko, A. K. Geim, C. Casiraghi, A. H. Castro Neto and K. S. Novoselov, *Science*, 2013, **340**, 1311–1314.

56  F. H. L. Koppens, T. Mueller, A. C. Ferrari, M. S. Vitiello and M. Polini, *Nat. Nanotechnol.*, 2014, **9**, 780–793.

57  C. Melzer and H. von Seggern, *Nat. Mater.*, 2010, **9**, 470–472.

58  G. Li, R. Zhu and Y. Yang, *Nat. Photonics*, 2012, **6**, 153–161.

**FIGURES**

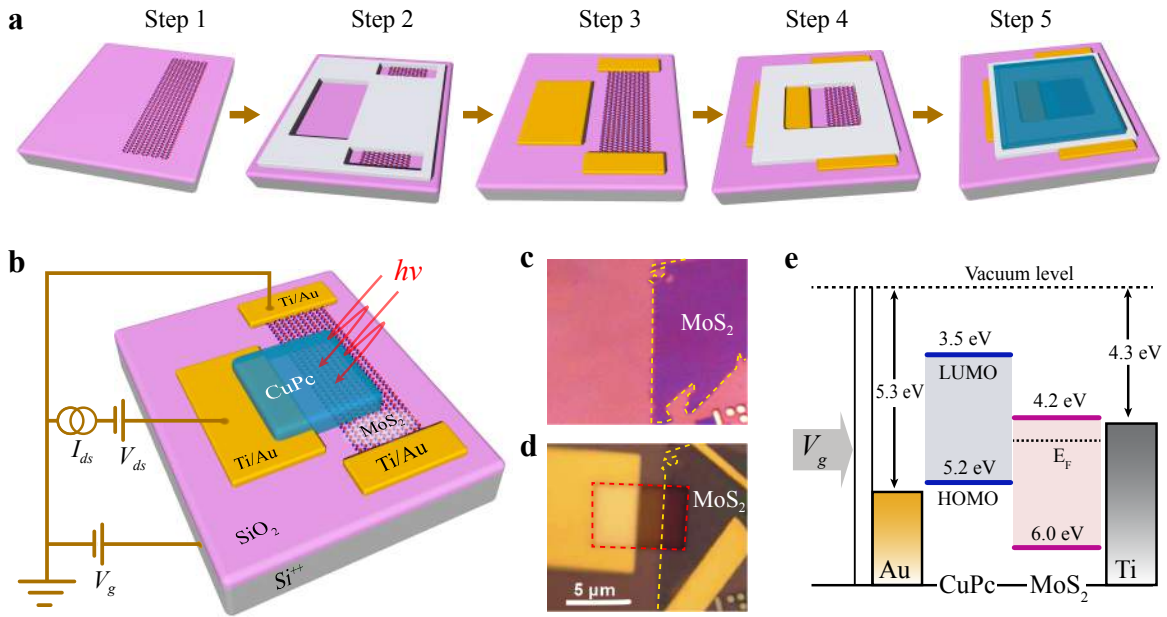

**Fig. 1** Fabrication of the CuPc-MoS$_2$ p-n junction devices. (a) Fabrication steps followed to produce the devices: exfoliation of a MoS$_2$ flake on a Si/SiO$_2$ substrate (step 1); lithography (step 2) and lift-off after Ti/Au (4/30 nm) deposition (step 3) to define the electrodes on SiO$_2$ and on MoS$_2$; lithography to define and open a window on PMMA between one electrode on SiO$_2$ and the MoS$_2$ flake (step 4); deposition of CuPc through a shadow mask to cover the area with the opened window on PMMA defined in the previous step with CuPc (step 5). (b) Sketch of a final device with its electrical connections. For clarity, only the CuPc within the p-n junction is depicted. (c) A bilayer MoS$_2$ flake after fabrication step 1 and (d) after fabrication step 5. (e) Scheme of the band structure of the p-n junction device with a monolayer MoS$_2$ in the rigid band structure. The work functions of the metals are also depicted. Energy levels for a bilayer MoS$_2$ would slightly change to account for the bandgap reduction from 1.8 to 1.6 eV. A dashed line schematically shows where the Fermi energy level is expected to be at zero gate. Note that the application of a positive (negative) $V_g$ would result in a modification of the $E_F$ upwards (downwards) in the MoS$_2$ as well as in a change in the available density of states at the HOMO level of the CuPc.

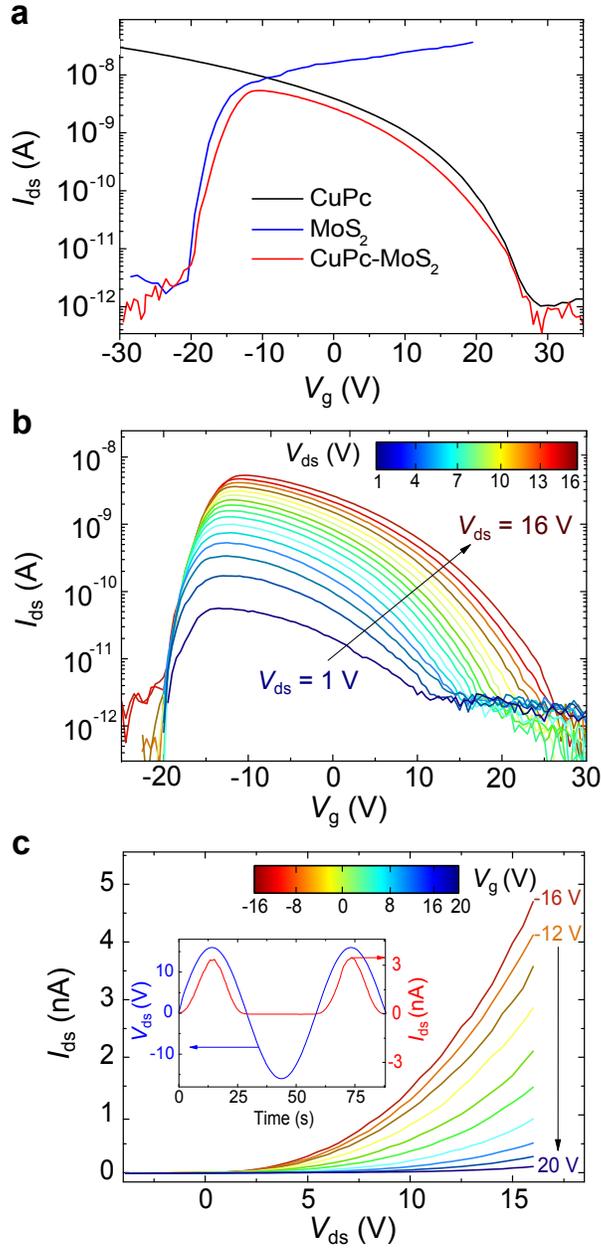

**Fig. 2** Gate-tunable diode characteristics of the CuPc-MoS$_2$ devices. (a) Transfer curves $I_{ds}(V_g)$ measured in a representative CuPc-MoS$_2$ device at a $V_{ds}$ = 16 V (red), in the MoS$_2$ bilayer of the same device at a $V_{ds}$ = 5 V (blue) and in a CuPc transistor produced on the same wafer at a $V_{ds}$ = 16 V (black). For direct comparison among the curves, the current measured in the MoS$_2$ was rescaled to account for the different $V_{ds}$ used. (b) Transfer curves of the CuPc-MoS$_2$ device measured at different $V_{ds}$ ranging from 1 to 16 V in steps of 1 V. (c) The source-drain current *vs.* voltage $I_{ds}(V_{ds})$ rectifying characteristics modulated by a gate voltage $V_g$ from -16 to 20 V in steps of 4 V. The inset shows the rectification properties of the p-n heterojunction under the application of an AC $V_{ds}$ at a fixed $V_g$ = -18 V. Note that all measurements were taken in dark conditions.

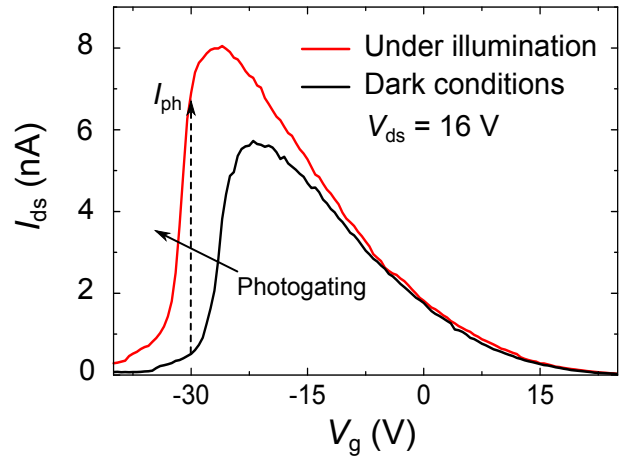

**Fig. 3** Comparison of the transfer curves measured in a CuPc-MoS$_2$ device in dark conditions ($I_d$-$V_g$; black) and under white light illumination ($I_l$-Vg; red) for $V_{ds}$ = 16 V (forward bias condition) showing photogating effect. The black dashed arrow schematically shows the amount of photocurrent ($I_{ph}$) generated at $V_g$ = -30 V. Notice that the ratio $I_l/I_d$ exceeds 10 at $V_g$ = -30 V.

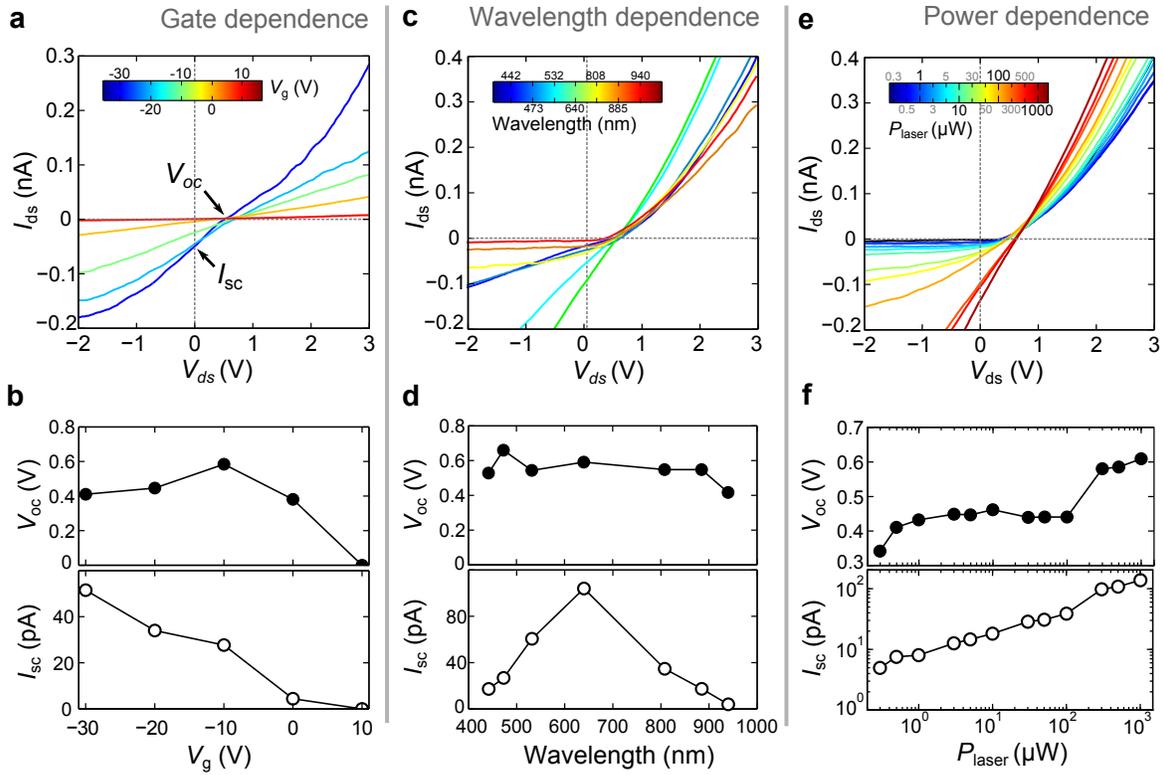

**Fig. 4** Photovoltaic effect of a CuPc-MoS$_2$ device: (a) and (b) as a function of the gate voltage (laser optical power $P_{laser}$ = 100 μW and λ = 640 nm); (c) and (d) as a function of the wavelength (laser optical power $P_{laser}$ = 500 μW and $V_g$ = - 20 V), and (e) and (f) as a function of the incident laser optical power ($V_g$ = -20 V and λ = 640 nm). Panels (b), (d) and (f) show the open circuit voltage $V_{oc}$ and short circuit current $I_{sc}$ extracted from the $I_{ds}(V_{ds})$ curves plotted in panels (a), (c) and (e), respectively.

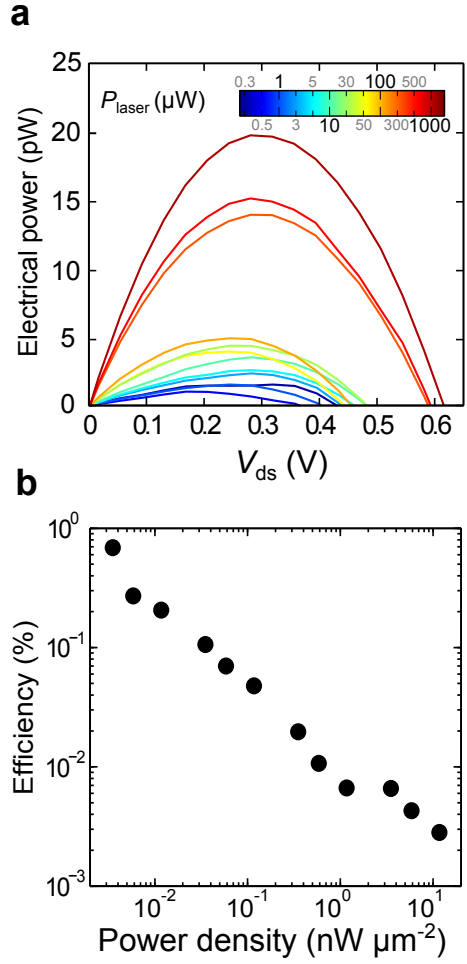

**Fig. 5** Electrical power generation in a CuPc-MoS$_2$ device. (a) Generated electrical power, $P_{el} = V_{ds} \cdot I_{ds}$, as a function of $V_{ds}$ for different incident optical powers, $P_{laser}$, and fixed wavelength λ = 640 nm. Data extracted from Fig. 4e. (b) Corresponding optical ($P_{opt}$) to electrical ($P_{el}$) power conversion efficiency extracted from the maximum electrical power $P_{el,m}$ obtained from data in panel a at each $P_{opt}$. To compute $P_{opt}$ from $P_{laser}$ both the laser spot size and the area of the junction are taken into account (see Materials and methods for details).

**Electronic Supplementary Information**

# Gate-tunable diode and photovoltaic effect in an organic-2D layered material p-n junction

Saül Vélez[a,†], David Ciudad[a,†], Joshua Island[b], Michele Buscema[b], Oihana Txoperena[a], Subir Parui[a], Gary A. Steele[b], Fèlix Casanova[a,c], Herre S. J. van der Zant[b], Andres Castellanos-Gomez[b,d,*], Luis E. Hueso[a,c,*]

[a] CIC nanoGUNE, 20018 Donostia-San Sebastian, Basque Country, Spain

[b] Kavli Institute of Nanoscience, Delft University of Technology, Lorentzweg 1, 2628 CJ Delft, The Netherlands

[c] IKERBASQUE, Basque Foundation for Science, 48011 Bilbao, Basque Country, Spain

[d] Instituto Madrileño de Estudios Avanzados en Nanociencia (IMDEA-Nanociencia), 28049 Madrid, Spain

[†] These authors have contributed equally to this work

* E-mail: (A.C.-G.) andres.castellanos@imdea.org, (L.E.H.) l.hueso@nanogune.eu

## S1. $I_{ds}$-$V_{ds}$ curves of the MoS$_2$ and the CuPc field effect transistors (FETs)

In our CuPc-MoS$_2$ devices, two metal electrodes were defined on the MoS$_2$ flake. The FET characteristics of the MoS$_2$ layer can be measured using these two electrodes as drain and source contacts, and a third contact on the Si$^{++}$ substrate as the gate (Fig. S1a shows a sketch of the electrical connections). CuPc FETs were also fabricated in the same wafer to characterize the electrical properties of the organic film. We used Ti (work function 4.3 eV) to contact the MoS$_2$ flake (electron affinity 4.2 eV) in order to prevent the formation of a large Schottky barrier. For the same reason, we used Au to contact the CuPc (Au work function is 5.3 eV and the HOMO level of the CuPc is at 5.2 eV). Fig. S1b and S1c show the $I_{ds}$-$V_{ds}$ characteristics measured at different $V_g$ values for two representative MoS$_2$ and CuPc FETs, respectively. Insets of Fig. S1b and S1c show a zoom of the $I_{ds}$-$V_{ds}$ curves in the small bias regime. The first linear trend is indicative of quasi-ohmic contact achieved in the metal-semiconductor junctions. A deviation from the linearity is observed at large $V_{ds}$ values in both cases (note that this is only achievable in truly metallic contacts).

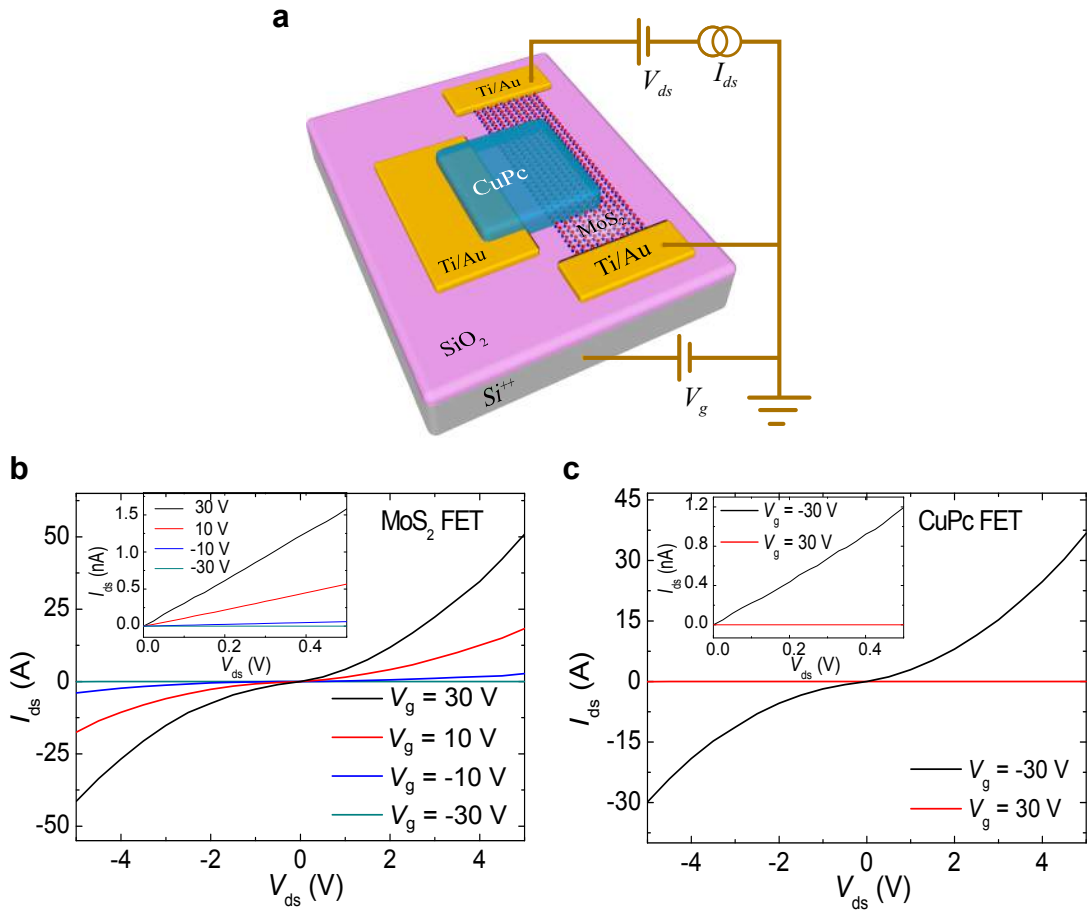

**Fig. S1** Characterization of the metal-semiconductor contacts. (a) Schematics of the contacts made in our CuPc-MoS$_2$ devices to characterize the FET characteristics of the MoS$_2$ flakes (contacts for a CuPc FET are not shown here). (b) and (c) $I_{ds}$-$V_{ds}$ curves of a representative MoS$_2$ and CuPc FETs, respectively, measured at different gate voltages. Insets in (b) and (c) show a zoom in the first quadrant of the $I_{ds}$-$V_{ds}$ curves. The quasi-linear behavior at small bias values indicates small Schottky barrier formation (quasi-ohmic contacts) at the metal-semiconductor contacts. A deviation from the linearity is observed at large $V_{ds}$ values.

## S2. Conduction of the CuPc-MoS$_2$ heterojunction

As a first approximation, the total resistance of the CuPc-MoS$_2$ heterostructure can be modeled as two resistances in series: one resistance due to the CuPc film and the other related to the MoS$_2$ flake, giving $R_H = R_{CuPc} + R_{MoS2}$. In our devices, we can measure/estimate the resistances of the CuPc and the MoS$_2$ by individually characterizing both the CuPc and the MoS$_2$ FETs, in addition to the CuPc-MoS$_2$ FET. We denote $R_H = V_H/I_H$ as the resistance of the CuPc-MoS$_2$ FET heterojunction; $R_{CuPc} = V_{CuPc}/I_{CuPc}$ for the CuPc FET; and $R_{MoS2} = V_{MoS2}/I_{MoS2}$ for the MoS$_2$ FET. If we fix the voltage of all devices to the same $V_{ds}$, the current passing through the heterojunction is given by $I_H = V_H/(R_{MoS2}+R_{CuPc})$. Using the relations given above one gets

$$I_H = (I_{CuPc} \cdot I_{MoS2})/(I_{CuPc}+I_{MoS2}). \tag{S1}$$

Notice that $I_{CuPc}$ and $I_{MoS2}$ are not the currents passing through the layers of the heterostructure when applying a voltage $V_{ds}$, but the currents measured in each individual FET when applying a voltage $V_{ds}$. This relation allows us to reproduce the experimental data $I_{ds}(V_g)$ obtained for a CuPc-MoS$_2$ FET (for instance, the red curve in Fig. 2a) from the measured individual FETs (see blue and black curves in Fig. 2a). Fig. S2 shows the transfer curves of both CuPc and MoS$_2$ FETs extracted from Fig. 2a together with the calculated transfer curve for the heterojunction using Equation S1 (note that the FET characteristics of the MoS$_2$ flake was scaled to account for the different $V_{ds}$ used). There is a good qualitative agreement between the experimental data and the calculated transfer curve (red straight line in Fig. 2a vs red squares plotted in Fig. S2). Notice that the exact solution (which is out of our scope here) might also account for other factors as for differences in the geometry for the conducting channels, screening effects in the CuPc layer when measuring the CuPc-MoS$_2$ devices, etc.

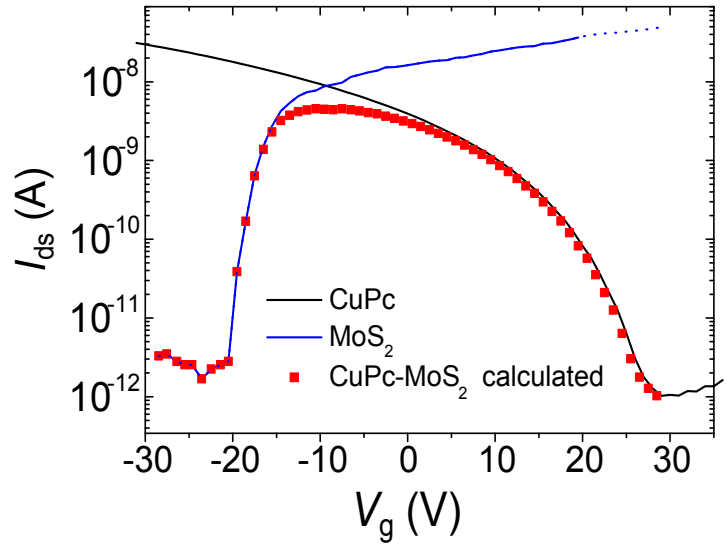

**Fig. S2** Calculated transfer curve of the CuPc-MoS$_2$ FET (red) from the experimental curves of the individual FETs (MoS$_2$-blue and CuPc-black) using Equation (S1). Original data extracted from Fig. 2a.

**S3. Electrical characterization of one of the monolayer-MoS$_2$-based CuPc-MoS$_2$ device**

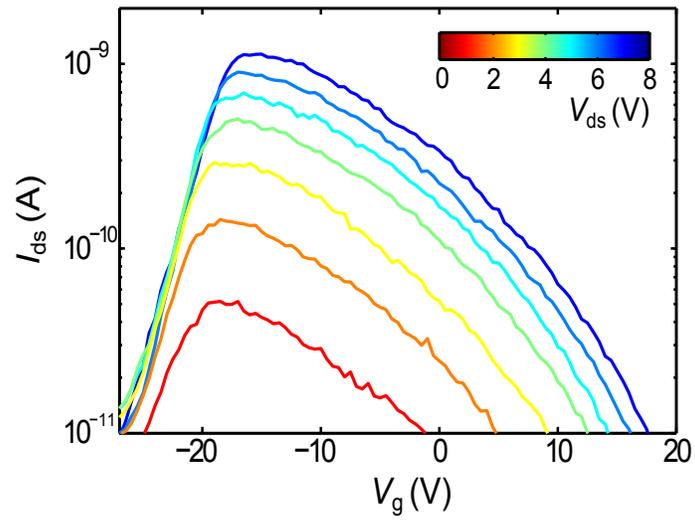

**Fig. S3** Transfer curve $I_{ds}$-$V_g$ of a representative CuPc-MoS$_2$ FET made of a monolayer MoS$_2$. These curves resemble the ones obtained with a bilayer-based CuPc-MoS$_2$ device (Fig. 2b). Similar results were obtained among different CuPc-MoS$_2$ p-n junction devices (for both monolayer- and bilayer-MoS$_2$-based devices).

## S4. Fits of the diode characteristics to the Shockley equation

In this section, we model the $I_{ds}$-$V_{ds}$ characteristics of the p-n junction for a representative CuPc-MoS$_2$ device with a modified form of the Shockley equation. In an ideal case, the relationship between the current $I_{ds}$ and the voltage bias $V_{ds}$ across a p-n diode is described by the Shockley model:

$$I_{ds} = I_s\left[\exp\left(\frac{V_{ds}}{nV_T}\right) - 1\right], \quad (S2)$$

where $I_s$ is the saturation current, $n$ is the ideality factor, $V_T = k_B T$ ($k_B$ is the Boltzmann constant in eV/K and $T$ is the temperature in K) is the thermal voltage. The ideality factor is related to the carrier recombination mechanisms at the p-n junction; $n = 1$ indicates that there is only band-to-band recombination of minority carriers, which is the ideal case.

A more realistic model should include current losses due to parasitic resistances in parallel ($R_p$, also called *shunt resistance*) and in series ($R_S$) with the junction. A schematic of the model circuit is presented in the inset of Fig. S4a. The series resistance $R_S$ models the voltage losses due to e.g. contact resistance or the resistance associated to long leads of high resistivity. The parallel resistance $R_p$ models additional carrier recombination mechanisms that drain current from the junction. The slope of the measured $I_{ds}$-$V_{ds}$ curves at $V_{ds} = 0$ V indicates a non-infinite $R_p$. To include these effects, we can rewrite Equation (S2) as

$$I_{ds} = I_s\left[\exp\left(\frac{V_{ds} - I_{ds}R_s}{nV_T}\right) - 1\right] + \frac{V_{ds} - I_{ds}R_s}{R_p}. \quad (S3)$$

An analytical expression can be obtained in the following form [S1]

$$I_{ds} = \frac{nV_T}{R_s}W\left[\frac{I_s R_s R_p}{nV_T(R_s + R_p)}\exp\left(\frac{R_p(V_{ds} - I_s R_s)}{nV_T(R_s + R_p)}\right)\right] + \frac{V_{ds} - I_s R_p}{R_s + R_p}, \quad (S4)$$

where $W$ is the Lambert $W$-function.

The fits of the measured $I_{ds}$-$V_{ds}$ curves in a CuPc-MoS$_2$ (bilayer based) device to the Equation (S4) are shown in Fig. S4. We summarized the extracted model parameters in Table S1.

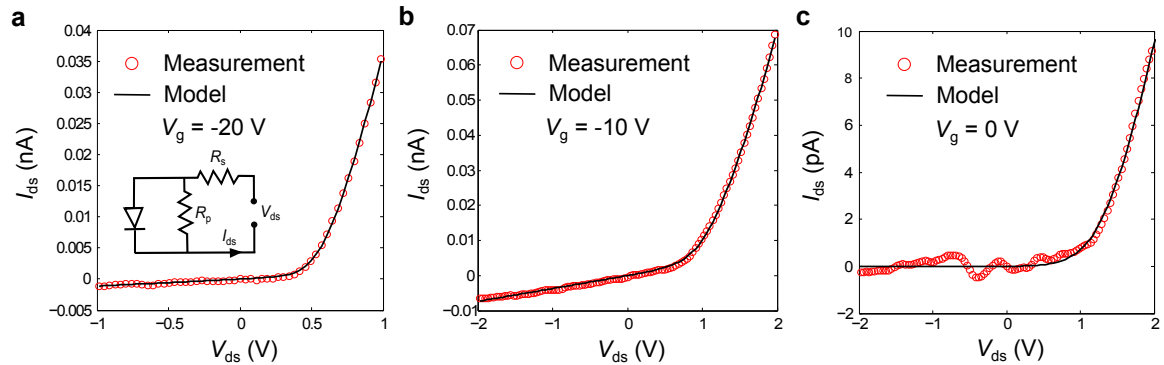

**Fig. S4** Fits of the $I_{ds}$-$V_{ds}$ diode-characteristics of the CuPc-MoS$_2$ junction to Equation (S4) for different gate voltages -20 V (a), -10 V (b) and 0 V (c). Inset of panel (a) shows the schematics of the model. Table S1 shows a summary of the extracted parameters.

|  | $V_g$ = -20 V | $V_g$ = -10 V | $V_g$ = 0 V |
|---|---|---|---|
| $I_S$ (µA) | 4.23 | 1.63 | 7.65 |
| $n$ | 2.87 | 3.97 | 8.25 |
| $R_S$ (GΩ) | 9.15 | 14.39 | 50.6 |
| $R_p$ (GΩ) | 874.1 | 239.1 | >1000 |

**Table S1** Summary of the main parameters extracted from the fittings shown in Fig. S4.

## S5. Photogating of the CuPc-MoS$_2$ devices

The photoresponse of the CuPc-MoS$_2$ heterostructure (Fig. 3a) can be modeled as a shift of the transfer curve of the MoS$_2$ layer towards (more) negative $V_g$ values (n-type doping). This shift can be explained via electron doping caused by the photogenerated charges in the CuPc layer, which might be captured in long-lived trap states at the CuPc-MoS$_2$ interface itself or close to the junction. This behavior is known as photogating, and already reported to occur for MoS$_2$ [S2].

To illustrate the effect, Fig. S5a shows the transfer curve of a MoS$_2$-CuPc FET (red) calculated from the transfer curves of a CuPc (black) and a MoS$_2$ (blue) FETs (corresponding data of Fig. S2; see Supplementary Section S2 for more details) obtained in dark conditions. The n-doping of the MoS$_2$ layer under illumination can be simulated by mathematically adding an offset to the MoS$_2$ FET curve (dashed blue line in Fig. S5). Using this simulated curve and the transfer characteristics of the CuPc FET, the transfer curve of the CuPc-MoS$_2$ under illumination is also calculated and included in Fig. S5 (dashed red curve). For sake of comparison, the transfer curves of the CuPc-MoS$_2$ device in dark conditions (straight black curve) and under illumination (dashed red curve) are plotted in Fig. S5b in the linear scale. The simulated photogating effect agrees quite well with the photoresponse shown in Fig. 3a. Note that the differences between the characteristics of this device and the one shown in Fig. 3a are attributed to different doping of the constituent layers.

One could also consider that photogating in the CuPc layer might also occur (in that case, the transfer curve would be moved towards positive values), but the sharper subthreshold characteristics of the MoS$_2$ layer makes n-doping of the MoS$_2$ layer dominant over p-doping of the CuPc, if there is any.

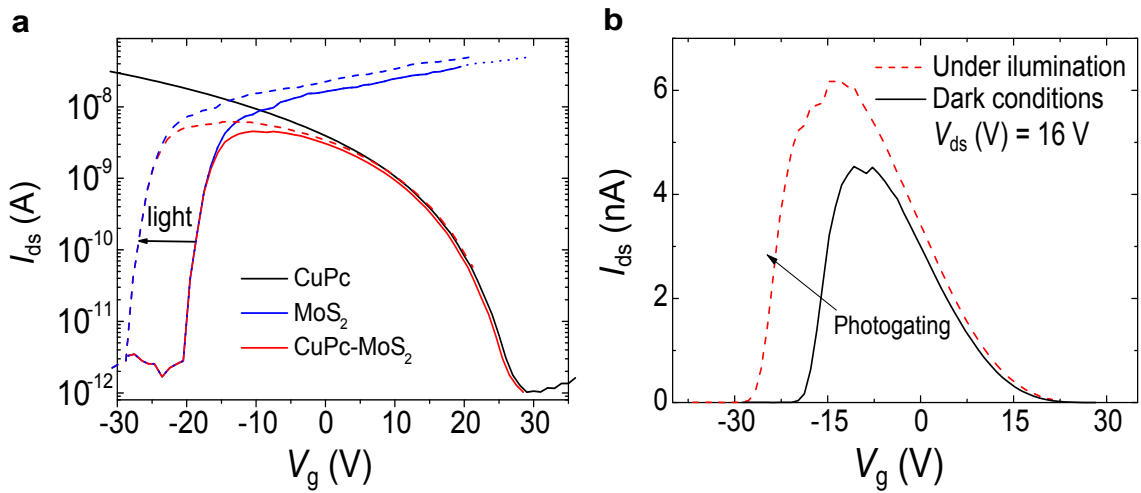

**Fig. S5** Simulated photogating of the CuPc-MoS$_2$ devices. (a) CuPc (black straight line) and MoS$_2$ (blue straight line) FETs. Data corresponding to the device explored in Fig. 2a. Calculated transfer curve for the CuPc-MoS$_2$ device (red straight line) as explained in Supplementary Section 2. The simulated photogating of the MoS$_2$ layer ($\Delta V_g$ = -8 V) is represented as a dashed blue line. Dashed red line shows the calculated CuPc-MoS$_2$ transfer curve characteristics considering the photogating effect. (b) Linear plot of the calculated CuPc-MoS$_2$ transfer curves for both dark conditions (blak straight line) and simulated under illumination (red dashed line). The curves shown here captures the behavior observed in Fig. 3a.

## S6. Gate dependence of the photoinduced current in the CuPc-MoS$_2$ p-n junction

The photoinduced current ($I_{ph}$) of the CuPc-MoS$_2$ p-n junction is calculated by subtracting the source-drain current measured in dark conditions ($I_d$) to the one measured under illumination ($I_l$): $I_{ph} = I_l - I_d$, at a given $V_{ds}$ and $V_g$. Fig. S6a schematically represents the generated $I_{ph}$ (indicated by a vertical dashed arrow) at a particular $V_g$ value ($V_g \sim -30$ V) measured at $V_{ds} = 16$ V. Fig. S6b shows a 2D colored plot of the $I_d$ measured as a function of both the $V_{ds}$ (x-axis) and $V_g$ (y-axis). By performing the same measurements under illumination $I_l(V_{ds}, V_g)$, the $I_{ph}(V_{ds}, V_g)$ 2D map is calculated and plotted in Fig. S6c. The main feature of the photoresponse is associated to the shift experienced in the $V_g$ value where the maximum of the $I_{ph}$ appears with respect to the $I_d$. This effect is explained via photogating (see Supplementary Section S5).

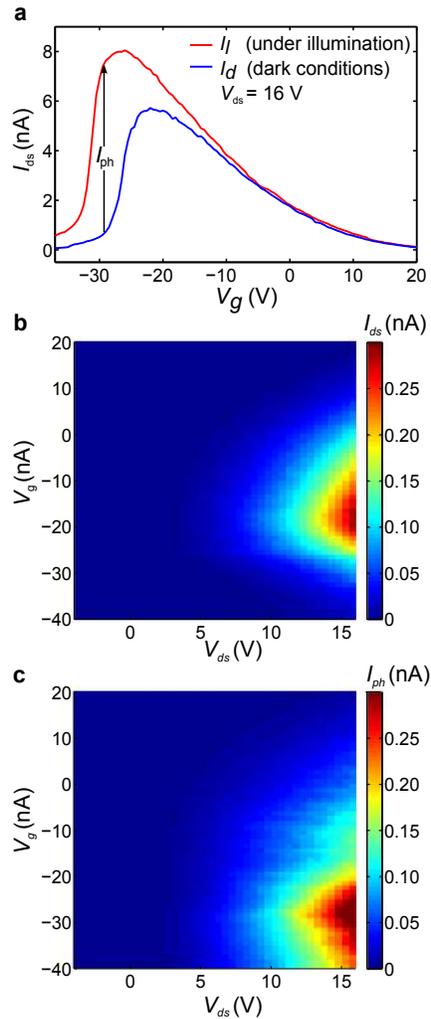

**Fig. S6** Gate dependence of the photoinduced current in the CuPc-MoS$_2$ junction. (a) $I_{ds}$-$V_g$ curves measured in dark conditions ($I_d$) and under illumination ($I_l$) at $V_{ds} = 16$ V. (b) 2D colored map of the current measured in dark conditions $I_d$. (c) Photoinduced current $I_{ph} = I_l - I_d$ as a function of $V_{ds}$ and $V_g$. The strong shift in $V_g$ observed between dark conditions and under illumination is explained via photogating of the MoS$_2$ layer. Under illumination, a maximum increase of the $I_{ds}$ of up to two orders of magnitude is measured with respect to dark conditions.

## S7. Dataset of the optoelectronics characterization of a CuPc-MoS$_2$ monolayer-based device

The photovoltaic effect in a monolayer-MoS$_2$-based CuPc-MoS$_2$ device has been explored as a function of the gate voltage (Fig. S7), wavelength (Fig. S8) and incident laser power (Fig. S9). Qualitatively, similar results compared to the bilayer CuPc-MoS$_2$ device (Fig. 4) are found. Quantitatively, slightly larger $V_{OC}$ voltages are obtained in this case, which might be attributed to a change in the band-gap between monolayer and bilayer MoS$_2$, resulting in a slightly different band-alignment with the CuPc film. We would like to comment here that during the characterization of the monolayer CuPc-MoS$_2$ we experienced some leakage problems, making the data measured in this case more noisy/less accurate than in the bilayer case.

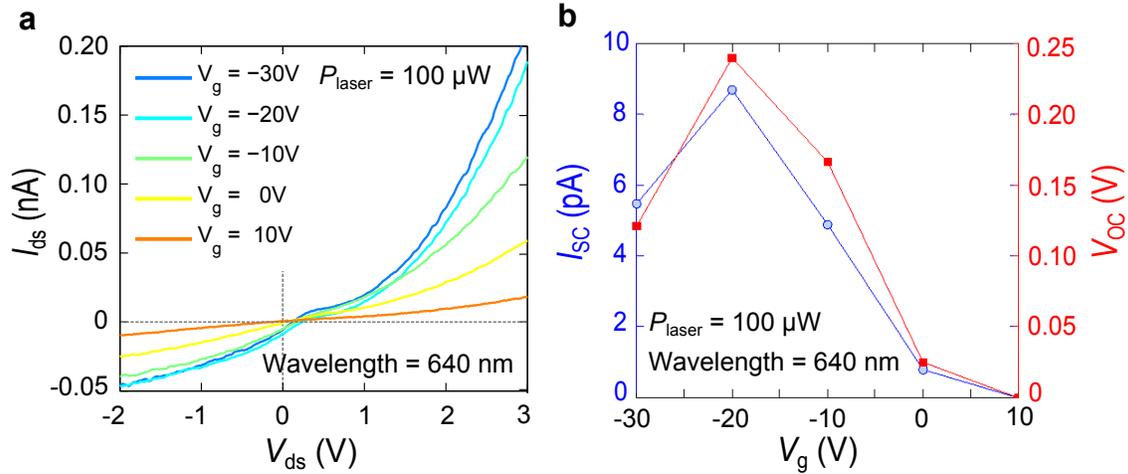

**Fig. S7** Gate dependence on the photovoltaic effect of a CuPc-MoS$_2$ monolayer-based device. (a) $I_{ds}$-$V_{ds}$ curves measured at different $V_g$ values under illumination with $P_{laser}$ = 100 mW and λ = 640 nm. (b) $I_{SC}$ and $V_{OC}$ values extracted from (a). The maxima in the $I_{SC}$ and $V_{OC}$ values are found around the region where the diode shows better ideality factors, and correlates with the maxima of the conductivity of the p-n junction.

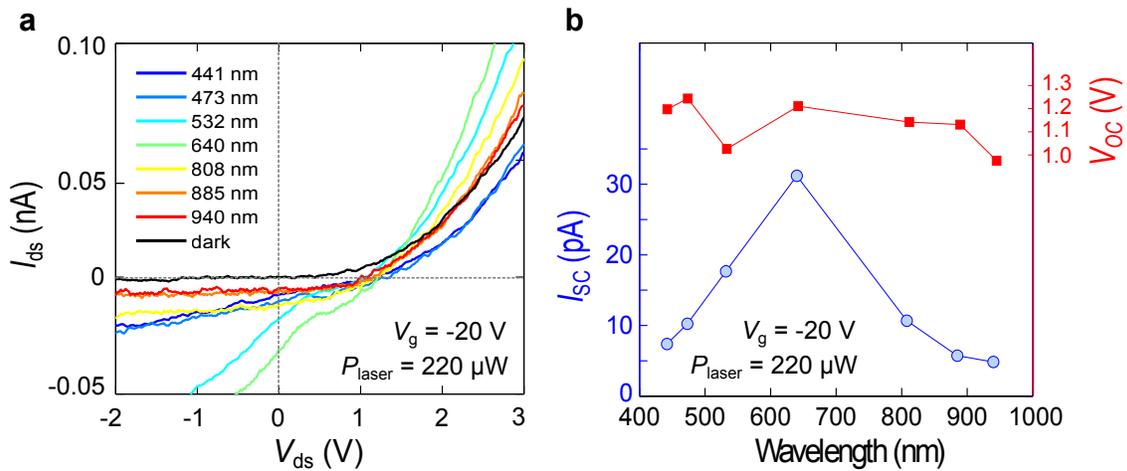

**Fig. S8** Wavelength dependence on the photovoltaic effect of a CuPc-MoS$_2$ monolayer-based device. (a) $I_{ds}$-$V_{ds}$ curves measured under illumination at different wavelengths. Measurements taken at fixed $V_g$ = -20 V and $P_{laser}$ = 220 µW. (b) $I_{SC}$ and $V_{OC}$ values extracted from (a). Fairly constant large $V_{OC}$ are obtained at all wavelengths whereas the $I_{SC}$ values follow the absorption edge of the CuPc layer.

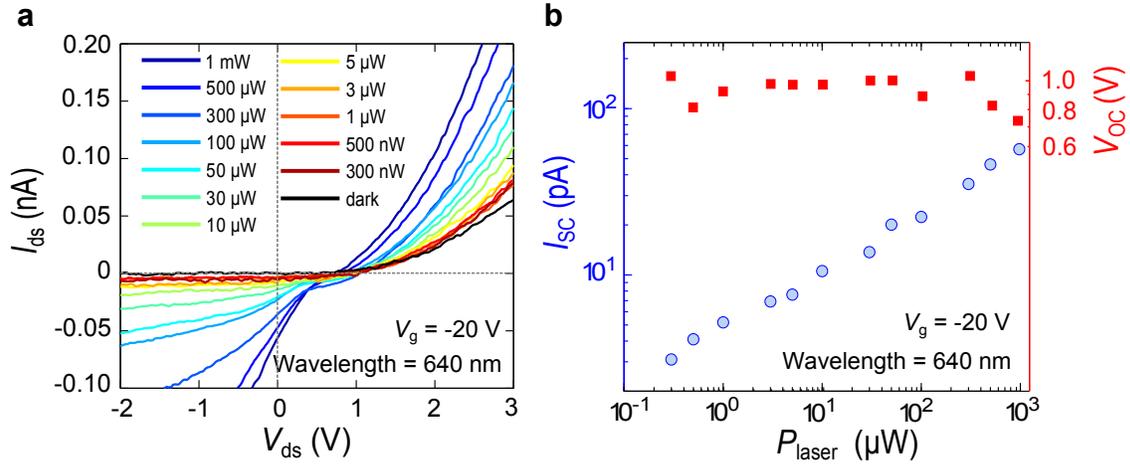

**Fig. S9** Power dependence on the photovoltaic effect of a CuPc-MoS$_2$ monolayer-based device. (a) $I_{ds}$-$V_{ds}$ curves measured at different optical powers and at fixed $V_g$ = -20 V and λ = 640 nm. (b) $I_{SC}$ and $V_{OC}$ values extracted from (a). A fairly constant $V_{OC}$ is obtained at all $P_{laser}$ values and $I_{SC}$ increasing with $P_{laser}$ following a power law.

The responsivity of the monolayer-MoS$_2$-based CuPc-MoS$_2$ device is tested for different incident optical powers and for $V_{ds}$ = -6 V, $V_g$ = -20 V and λ = 640 nm. Similar values to the ones obtained for the bilayer-based device had been found, with computed EQE~10% at small incident optical powers.

**S8. Wavelength dependence of the photovoltaic effect in the CuPc-MoS$_2$ devices**

The $I_{sc}$ current generated in our CuPc-MoS$_2$ devices as a function of the wavelength (Fig 4d) resembles the light absorbance spectra of CuPc rather than of MoS$_2$ (see References S3 and S4 for instance). The absorbance of CuPc has a minimum at a wavelength of 450 nm and monotonically increases up to a maximum at ~625 nm. It has a secondary maximum at ~700 nm. For larger frequencies the absorbance is reduced and is almost negligible for wavelengths larger than 850 nm. The correlation between the responsivity and the absorbance of CuPc suggest that the photocurrent in the CuPc-MoS$_2$ junction is mainly originated in the organic layer, which is in agreement with all other features observed (see main text).

**References**

(S1)  A. Ortiz-Conde, F. J. García Sánchez and J. Muci, *Solid-State electronics,* 2000, **44**, 1861-64.

(S2)  M. M. Furchi, A. Pospischil, F. Libisch, J. Burgdörfer and T. Mueller, *Nano Lett*ers, 2014, **14**, 4785-4791.

(S3) Q. Li, J. Yu, Y. Zhang, N. Wang and Y. Jiang, *Front. Energy*, 2012, **6**, 179–183.

(S4) A. Splendiani, L. Sun, Y. Zhang, T. Li, J. Kim, C.-Y. Chim, G. Galli, F. Wang, *Nano Lett.* 2010, **10**, 1271.